\documentclass[12pt]{article}
\pdfoutput=1
\usepackage{amsfonts}
\usepackage{amssymb}
\usepackage{url}
\usepackage{hyperref}
\usepackage{simpler-wick}
\usepackage{bbold}
\usepackage{slashed}
\usepackage{graphicx}
\usepackage{pgfplots}
\usepackage{color}
\usepackage{mathrsfs}
\usepackage{fancybox}
\usepackage{lipsum}
\usepackage{eurosym}
\usepackage{tcolorbox}
\usepackage{lmodern}
\usepackage{tikz}
\usepackage{tikz,pgfplots}
\usepackage{pgfplots}
\usepackage{amsmath,amsthm,amssymb}
\pgfplotsset{compat=1.10}
\usepgfplotslibrary{fillbetween}
\usepackage{feynmp}
\usepackage{slashed}
\usepackage{graphicx}
\usepackage{epstopdf}
\usepackage{subfigure}
\usepackage{color}
\usetikzlibrary{decorations.pathmorphing}
\tikzset{snake it/.style={decorate, decoration=snake}}
\usetikzlibrary{shapes}
\usepackage{empheq}
\usepackage{lmodern}
\usepackage{anyfontsize}
\usepackage{t1enc}

\def\({\left (}
\def\){\right )}
\def\[{\left [}
\def\]{\right ]}
\def\d{\mathrm{d}}

\numberwithin{equation}{section}

\usepackage{environ}
\NewEnviron{myequation}{%
\begin{eqnarray}
\scalebox{1.05}{$\BODY$}
\end{eqnarray}
}

\DeclareMathOperator*{\Motimes}{\raisebox{0.2ex}{\scalebox{1}{$\bigotimes$}}}

\newcommand{\ba}{\begin{equation}}
\newcommand{\ea}{\end{equation}}
 
\newcommand{\bea}{\begin{eqnarray}}

\newcommand{\bbbr}{\!\begin{array}}
\newcommand{\earr}{\end{array}\!}
\newcommand{\lb}{{\langle}}
\newcommand{\rb}{{\rangle}}

\newcommand{\be}{\begin{equation}}
\newcommand{\ee}{\end{equation}}

\def\lsim{\mathrel{\rlap{\lower4pt\hbox{\hskip1pt$\sim$}}
     \raise1pt\hbox{$<$}}}         
\def\gsim{\mathrel{\rlap{\lower4pt\hbox{\hskip1pt$\sim$}}
     \raise1pt\hbox{$>$}}}         

\begin{document}
\newcommand{\CC}{{\mbox{\textbf {\textit C}}}}
\newcommand{\RR}{{\mbox{\textbf {\textit R}}}}
\newcommand{\PP}{{\mbox{\textbf {\textit P}}}}
\newcommand{\nnn}{n}
\def\smallTFD{{\rm TFD}}
\def\d{{\partial}}
\def\n{{\bf \widehat n}}
\def\k{{\bf k}}
\def\changemargin#1#2{\list{}{\rightmargin#2\leftmargin#1}\item[]}
\let\endchangemargin=\endlist 
\def\smalll{\footnotesize}
\def\aaA{{{}_{\! \mbox{\scriptsize  $A$}}}}

\def\cC{\mbox{\textit{\textbf{C}\!\,}}}
\def\bBB{{{}_{\! \mbox{\scriptsize  $B$}}}}
\def\ccC{{{}_{\nspc \mbox{\scriptsize  $C$}}}}
\def\abAB{{{}_{\! \mbox{\scriptsize  $A\nspc\nspc B$}}}}
\def\bB{{\mbox{\scriptsize $b$}}}
\def\bbB{{\mbox{\scriptsize  $b$}}}

\def\ccc{{{\!}_{\mbox{\small {$c$}}}}}

\def\mn{{{}_{\raisebox{-.5pt}{\scriptsize $\mom \nom$}}}}
\def\nm{{{}_{\raisebox{-.5pt}{\scriptsize $\nom \mom$}}}}

\def\sS{\uU}
\def\lbl{{\raisebox{-.5pt}{\large$($}}}
\def\rbl{{\raisebox{-.5pt}{\large$)$}}}

\def\bst{{\raisebox{-.5pt}{\large$|$}}}

\def\bigll{\bigl}
\def\bigrr{\bigr}

\def\mathbi#1{\textbf{\em #1}} 
\def\som{{ \textit{\textbf s}}} 
\def\tom{{ \textit{\textbf t}}} 
\def\nom{{ \textit{\textbf n}}} 
\def\mom{{ \textit{\textbf m}}} 
\def\kom{{ \textit{\textbf k}}}  
\def\nomt{{ \textit{\textbf n}}}  
\def\momt{{ \textit{\textbf m}}}  
\def\komt{{ \textit{\textbf k}}} 
\def\iii{i}\def\plus{\raisebox{.5pt}{\tiny$+$\smpc}}

\def\spc{\hspace{1pt}}

\def\nspc{{\hspace{-2pt}}}
\def\ff{\mathsf\smpc f\smpc} 
\def\fff{\mbox{Y}}
\def\ww{{\mathsf w}}
\def\smpc{{\hspace{.5pt}}}

\def\zz{{\spc \mathsf z}}
\def\xx{{\mathsf x\smpc}}
\def\xxi{\mbox{\footnotesize \spc ${e^{S_0}}$}}
\def\jj{{\mathsf j}}
 \addtolength{\baselineskip}{.1mm}

\def\calO{{b}}
\def\be{\begin{equation}}
\def\ee{\end{equation}}




\def\bfU{\mbox{\textbf{\textit U}}}
\def\bfR{\mbox{\textbf{\textit R}}} 
\def\bfC{\mbox{\textbf{\textit C}}} 
\def\bfT{\mbox{\textbf{\textit T}}} 
\def\bbbi{{\, {{i}}\, }}
\def\nnn{n}

\def\spcm{\hspace{.25pt}}
\def\mathbi#1{\textbf{\em #1}} 
\def\som{{ \textit{\textbf s}}} 
\def\tom{{ \textit{\textbf t}}} 
\def\nnn{n} 
\def\mom{m} 
\def\la{\langle}
\def\bea{\begin{eqnarray}}
\def\eea{\end{eqnarray}}
\def\is{\!  & \!  = \!  &  \!}
\def\ra{\rangle}
\def\half{{\textstyle{\frac 12}}}
\def\cL{{\cal L}}
\def\halfi{{\textstyle{\frac i 2}}}
\def\ba{\bea}
\def\ea{\eea}
\def\lb{\langle}
\def\rb{\rangle}
\newcommand{\rep}[1]{\mathbf{#1}}

\def\uU{\bfU}
\def\be{\bea}
\def\ee{\eea}
\def\delbar{\overline{\partial}}
\def\ra{{\mbox{\large $\rangle$}}}
\def\la{{\mbox{\large $\langle$}}}
\def\ccdot{\!\spc\cdot\!\spc}
\def\nspc{\!\spc}
\def\tr{{\rm tr}}
\def\li{|\spc}
\def\ri{|\spc}
\def\OA{O_{\!\spc A}} 
\def\OC{O_C} 
\def\OB{O_B} 
\def\tMD{\mbox{\fontsize{7}{7}$\rm\smpc TMD$}}

\def\gtMD{\mbox{\fontsize{7}{7}$\rm\smpc GTMD$}}

\def\tFD{\mbox{\fontsize{7}{7}$\rm\smpc TFD$}}
\def\Oomega{\mbox{\fontsize{7}{7}$\rm\smpc \Omega$}}

\def\hf{\textstyle \frac 1 2}

\def\smpt{\hspace{.9pt}}
\def\bfcdot{\raisebox{-1.5pt}{\bf \LARGE $\spc \cdot\spc $}}
\def\spc{\hspace{1pt}}
\def\is{\! &  \! = \! & \!}
\def\d{{\partial}}
\def\n{{\bf \widehat n}}
\def\k{{\bf k}}
\def\GO{{\cal O}}
\def\rmAD{{\mathsf{Rad}}} 
\def\bbb{}
\def\pp{{\mbox{\tiny$+$}}}
\def\mm{{\mbox{\tiny$-$}}}
\def\ppp{\mbox{\fontsize{9}{8}$p$}}
\def\qqq{\mbox{\fontsize{9}{8}$q$}}
\def\TFD{\mbox{\fontsize{11}{13}$\rm\smpc TFD$}}
\setcounter{tocdepth}{2}
\renewcommand{\Large}{\large}
\addtolength{\baselineskip}{0.25mm}
\addtolength{\parskip}{.8mm}
\def\calO{{b}}
\def\be{\begin{equation}}
\def\ee{\end{equation}}
\def\half{\raisebox{-1pt}{\large $\frac 1 2$}}
\addtolength{\abovedisplayskip}{1.5mm}
\addtolength{\belowdisplayskip}{1.5mm}
\def\zzz{{\mbox{\large $z$\nspc\smpc}}}
\begin{titlepage}

\setcounter{page}{1} \baselineskip=15.5pt \thispagestyle{empty}

\vfil

${}$
\vspace{1cm}

\begin{center}
\def\thefootnote{\fnsymbol{footnote}}
\begin{changemargin}{0.05cm}{0.05cm} 
\begin{center}
{\Large \bf  Deconstructing the Wormhole: \\[5mm]
Factorization, Entanglement and Decoherence}

\end{center} 
\end{changemargin}

~\\[1cm]
{Herman Verlinde\footnote{\href{mailto:verlinde@princeton.edu}{\protect\path{verlinde@princeton.edu}}}}
\\[0.3cm]

{\normalsize { \sl Physics Department,  
Princeton University, Princeton, NJ 08544, USA}} \\[3mm]

\end{center}

\begin{changemargin}{01cm}{1cm} 
{\small  \noindent 
\begin{center} 
\textbf{Abstract}\\[4mm]
\parbox{15 cm}{We study the role of ensemble averaging in holography by exploring the relation between the universal late time behavior of the spectral form factor and the second R\'enyi entropy of a thermal mixed state of the doubled system. Both quantities receive contributions from wormhole saddle-points: in the former case they lead to non-factorization while in the latter context they quantify decoherence due to the interaction with an environment. Our reasoning indicates that the space-time continuity responsible for non-factorization and space-time continuity through entanglement are in direct competition with each other. In an accompanying paper \cite{HV:2021}, we examine this dual relationship in a general class of 1D quantum systems with the help of a simple geometric path integral prescription.  }

\end{center} }

\end{changemargin}
 \vspace{0.3cm}
\vfil
\begin{flushleft}

\today
\end{flushleft}

\end{titlepage}

\flushbottom

\newpage
\tableofcontents
\thispagestyle{empty}
\newpage

\setcounter{footnote}{0}
\setcounter{page}{1}
\section{Introduction: factorization and purity}

\vspace{-1mm}

Holographic CFTs are chaotic quantum systems. Common diagnostics of quantum chaos rely on the property that the underlying energy spectrum tends to be both dense and random. The randomness is a reflection of the scrambling dynamics: any coherent pattern in the quantum state gets washed out due to dephasing. 
A quantitative measure of dephasing  is the spectral form factor ${Z}(\beta\! + \nspc it){Z}(\beta\! - \nspc it)$ with $Z(\beta) \spc = \spc \sum_{i}\,  e^{-\beta E_i}$ the thermal partition function. Assuming the energy spectrum obeys random matrix statistics,  the form factor exhibits universal behavior as a function of time $t$ \cite{Cotler:2016}-\cite{Johnson:2019}.  The slope is a smooth universal feature that reflects the initial decay of phase coherence.  The ramp and plateau, on other hand, exhibit non-universal wiggles. Upon averaging, however, either by summing over a suitable ensemble of CFTs or over many time instances, the ramp and plateau smooth out and become universal. The `dip' is the transition between the slope and the ramp. 
 
 We will be interested in the late time behavior of the averaged spectral form factor 
\bea
\label{zzlim}
\bigl\langle Z(\beta)^2 \bigr\rangle \! & \!  \equiv\! & \! \lim_{\ t \to\, \tau_\infty} \ \bigl \langle{Z}(\beta\! + \! it){Z}(\beta\! - \! it)
\bigr\rangle
\eea
 as a function of the late time $\tau_\infty$. We focus on two regions: the `dip' and the `plateau'. Our aim is to clarify and quantify the role of ensemble averaging in both regimes by making use of the relationship between the spectral form factor and the second R\'enyi entropy of the time evolved thermo-field double state. For both quantities, the holographic prescription includes wormhole saddle points \cite{Saad:2018}\cite{Almheiri:2019b}. In the former case they lead to non-factorizing contributions and in the latter case they quantify the decoherence process due to the interaction with an environment. In an accompanying paper \cite{HV:2021}, we examine this relationship in simple 1D quantum systems by means of a natural geometric path integral prescription. Here we focus on some general lessons that follow from this correspondence in the context of holography.

Holography identifies the thermal partition function $Z(\beta)$ of a 2D CFT with the gravitational path integral $Z(D)$ associated with a space-time region with the topology of a disk~$D$. The disk represents the longitudinal $(r,t)$ plane of a euclidean black hole with
the horizon at its center point and  the thermal circle as its boundary $\partial\nspc D$.
The non-averaged spectral form factor is the factorized partition function of two decoupled systems. The dual geometry therefore also factorizes. 

The averaged spectral form factor, on the other hand, does not factorize.  This non-factorization property hints at a holographic interpretation in terms of wormhole space-times that connect the two sides \cite{maldamaoz}-\cite{DonHenry}.  To leading order in the topological expansion,  the two boundaries are connected via a double trumpet geometry \cite{Saad:2018}
\bea
\label{sffsplit}
\bigl\langle{Z}(\beta)^2 \bigr\rangle_{\rm \! conn}  \!\! & \simeq &\! Z(\Sigma_2)  \qquad
\qquad \ \Sigma_2 \; = \; \raisebox{-17pt}{$$\includegraphics[width=2.25cm, height=1.4cm]{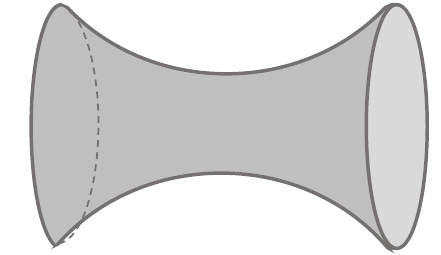}$$}
\eea
This term provides the dominant contribution near the dip. It represents the leading $1/N$ correction to the expectation value $\langle Z(\beta)^2\rangle = \langle \tr(e^{-\beta H})^2\rangle$ in a random matrix ensemble of Hamiltonians $H$ acting on a Hilbert space of dimension $N \sim e^{S_0}$. Via the usual 't Hooft large $N$ counting, this double trumpet contribution is suppressed relative to the factorized product of two thermal partition functions $Z(\beta)^2$ by an overall factor 
\bea
\label{zzdip}
\frac{\la Z(\beta)^2 \ra}{Z(\beta)^2} \, = \, \frac{ {Z(\Sigma_2)}}{Z(D)^2} \! &  \sim & \! e^{-2 S_0}  \qquad \qquad  \,  ({\rm near\ dip})
\eea
In dilaton gravity, $S_0$ represents the constant mode of the dilaton field. The factor of 2 in the exponent equals the difference between the Euler character of $\Sigma_2$ and a pair of disks $D$. 

As its name suggests, the dip is where the spectral form factor assumes its minimal value. Upon traversing the ramp, the average spectral form factor approaches the plateau value \cite{Saad:2018}
\bea
\label{zzplat}
\ \frac{\la Z(\beta)^2 \ra}{Z(\beta)^2} \, =\,  \frac{ Z(2\beta)}{Z(\beta)^2} \! & \sim & \! e^{- S_0}\qquad \qquad ({\rm plateau})
\eea
The numerator equals the time average of the spectral form factor\footnote{Assuming the energy spectrum has no exact degeneracies, one readily verifies that $\overline{\! |Z(\beta\!+\!it)|^2} =  Z(2\beta)$. Here $\overline{f(t)} = \raisebox{-2pt}{${\raisebox{-1pt}{\footnotesize$\lim$}}\atop{\raisebox{1.5pt}{\scriptsize$\tau \to \infty$}}$} \mbox{\Large $\int_{\raisebox{-1.5pt}{\scriptsize \nspc -\!\! $\tau$}}^{\raisebox{2pt}{\scriptsize $\tau$}}$} dt \, f(t)$ denotes the exact time average. }

Our aim in this note will be to clarify the different quantitative behavior \eqref{zzdip}-\eqref{zzplat} near the dip and the plateau region both from a quantum statistical and a holographic perspective. To this end, we will use the relation between the spectral form factor and the second R\'enyi entropy of the thermofield double to reinterpret the time dependence of the former in terms of the decoherence due to the interaction with an environment of the latter. As we will see, the two behaviors  \eqref{zzdip}-\eqref{zzplat}  of the spectral form factor are reflective of two different decoherence processes: decorrelation (near dip) and dephasing (plateau). 

Wormhole saddle points play a key role in the recent gravitational computations of the Page curve of an evaporating black hole \cite{replica1}-\cite{Almheiri:2019b}. At first sight, the replica wormholes that appear in the computation of the black hole R\'enyi entropy look different from the wormholes that give rise to the connected contributions to the spectral form factor. However, as we will now argue, the two types of wormholes are closely related.

Let us introduce the time evolved thermo-field double state
\bea
\label{tfdalpha}
 |{\TFD}(t) \rangle \is \sum_i \, e^{\mbox{\scriptsize$-(\frac{\beta}{2} +it)E_i$}}\spc |\, i \,  \rangle_L |\, \bar{i} \, \rangle_R 
\eea
Here $ |\, i \,  \rangle$ denotes an orthonormal basis of energy eigenstates with energie $E_i$ and $|\, \bar{i} \, \rangle = \hat{T} | \, i \, \rangle$ with $\hat{T}$ an anti-unitary involution that commutes with the Hamiltonian. 

Denoting the TFD density matrix by $\rho_{\tFD}(t) = |{\TFD}(t) \rangle \langle {\TFD}(t) |$
we can express the spectral form factor as
\bea
|{Z}(\beta\! + \nspc it_{12})|^2 
\is {Z(\beta)^2} \, \tr\raisebox{-.5pt}{\large$($}\rho_{\tFD}(t_1) \rho_{\tFD}(t_2)\raisebox{-.5pt}{\large$)$}
\eea
with $t_{12} = t_1-t_2$. 
The factorization property of the spectral form factor is directly linked to the purity of the time-evolved thermofield double state. Given this equality, it is natural to try to extend the correspondence to the averaged quantities.

With this motivation, we will look for a relationship between the late time limit \eqref{zzlim} of the averaged spectral form factor and the 2nd R\'enyi entropy of a suitable mixed state $\rho_{\tMD}$ 
\bea
\label{zztmd}
\tr\raisebox{-1.25pt}{\large $($}\rho_{\tMD}^2\raisebox{-1.25pt}{\large $)$} \is \frac{\la Z(\beta)^2 \ra}{Z(\beta)^2} 
\eea
called the {\it thermo-mixed double}  in \cite{HV:2020}\cite{Takayanagi:2020}. We will refer to the above relation the {\it replica Ansatz} for the averaged spectral form factor.

At very late times, both sides in \eqref{zztmd} approach the constant plateau value \eqref{zzplat}
\bea
\label{zzplatt}
\qquad \quad \tr\raisebox{-1pt}{\large $($}\rho_{\tMD}^2\raisebox{-1pt}{\large $)$} \is \frac{Z(2\beta)}{Z(\beta)^2}  \qquad \qquad ({\rm plateau})
\eea
Equation \eqref{zztmd}-\eqref{zzplatt} then has an obvious solution given by the time average of the TFD state
 \bea
 \label{latetmd}
\qquad  \rho_{\tMD} \is 
\overline{\nspc \rho_{\tFD}(t)\!} 
\; = \, \frac{1}{Z(\beta)\nspc}\,  \sum_i \, {e^{-\beta  E_i}} \spc |\spc i\spc \rangle_L\langle\spc i \spc |\otimes |\, \bar{i}\, \rangle_R\langle\, \bar{i}\, |  
 \eea
 This  late time TMD state exhibits perfect classical correlation between the energy levels on the two sides.   Note that the von Neumann entropy of \eqref{latetmd} equals the thermal entropy of the one-sided system. From now on we will refer to the state \eqref{latetmd} as the `old' TMD state.

In this note we will introduce a suitable generalization of $\rho_{\tMD}$ that will allow for a weak form of time dependence such that the replica Ansatz \eqref{zztmd} can be extended over the full late rime regime between the dip and the plateau. We are particularly interested in the dip region where the spectral form factor receives its dominant contribution from the connected double trumpet geometry $\Sigma_2$ depicted in \eqref{sffsplit}. Hence, via the Ansatz \eqref{zztmd}, the TMD state in this regime is defined through the relation
\bea
\label{tmddipp}
\qquad \quad \tr(\rho_{\tMD}^2) \!\is\!  \frac{ {Z(\Sigma_2)}}{Z(D)^2}
 \qquad \qquad  \,  ({\rm near\ dip}).
\eea

We can interpret the relation \eqref{tmddipp} geometrically as a gluing procedure for producing the double trumpet from two `janus pacman' geometries
\bea
\label{tmdglue}
{\includegraphics[scale=.6]{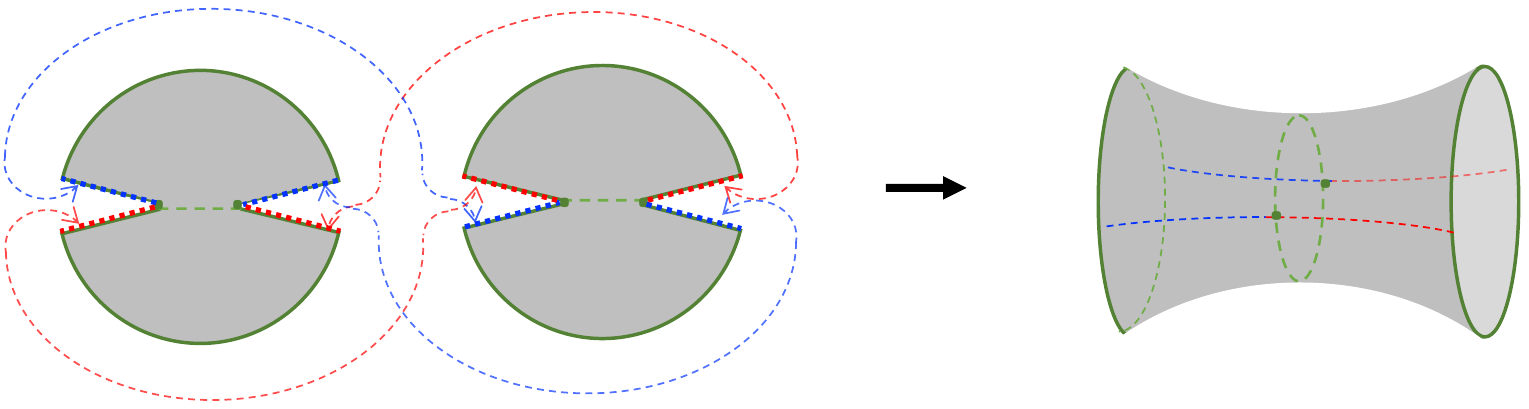}}
\eea
Conversely, we can read this figure from right to left as a deconstruction of the wormhole geometry: we cut the double trumpet open along the two horizontal dashed lines indicated on the right, and then fold each half to match the shapes on the left. 

This tells us that the thermal mixed double in this regime can be thought of as the result of performing the gravitational path integral over the janus pacman geometry. Schematically 
\bea
\rho_{\tMD} \! \is   \frac{Z(\Sigma_{\tMD})}{Z(\beta)} \qquad \qquad Z(\Sigma_{\tMD}) \, =\; \raisebox{-6.5mm}{\includegraphics[scale=.5]{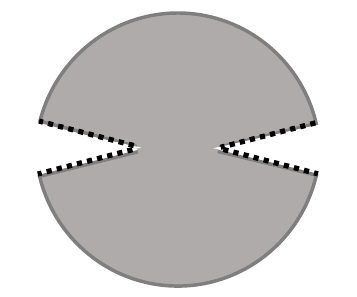}}
\eea
The two openings on both sides represent the {\small $L$ }and {\small $R$} ket- and bra-states. The `Island region' connecting the upper and lower half disk indicates that this defines a mixed state. Via the gluing \eqref{tmdglue}, the Islands combine to create the geometric bridge that connect the two sides of the double trumpet geometry. Note that this geometric construction places the bra- and ket states to opposite sides of the wormhole. It thus relates the non-factorization of the spectral form factor and the non-purity of the TMD state. This hints that the space-time connectedness through entanglement of the TFD state and the space-time connectedness responsible for non-factorization of the spectral form factor are in competition with each other. Our goal in the following is to clarify this dual relationship and make it more quantitative.

\bigskip

\section{Thermal mixed double, old and new}
\vspace{-.5mm}

In the previous section we introduced two versions of the thermal mixed double state. The `old'  TMD state  \eqref{latetmd} describes the fully dephased limit of the thermo-field double. Its second R\'eniy entropy coincides with the plateau value of the spectral form factor. While this plateau value is easily understood from a quantum statistical perspective, it is harder to interpret from the gravitational point of view. Indeed, for JT gravity, investigations of its matrix model dual indicate that understanding the plateau  involves non-perturbative physics from the point of view of the topological expansion of the gravity theory \cite{Saad:2019}\cite{Johnson:2019}.

The second version of the TMD state is defined through its relationship  \eqref{tmddipp} with the double trumpet geometry. This `new' thermal mixed double' is easier to understand geometrically, but less transparent from a microscopic point of view. We like to fill in this gap by writing a simple generalized TMD Ansatz that captures the required quantum statistical properties. The main feature we will look to explain is that the second R\'eniy entropy scales as $e^{-2S_0}$. A second hint that we will make use of is that the {\small $L$} and {\small $R$} side are connected via half a thermal circle. This indicates that, just as for the TFD and old TMD state \eqref{latetmd}, the {\small $L$} and {\small $R$} sides always have the same energy.

We infer that both versions of the TMD state can be written in the following form \cite{HV:2020}\cite{Takayanagi:2020}
\bea
\label{tmdsum}
\rho_{\tMD} \is \frac{1}{N} \, \sum_\alpha \; U_\alpha\, \rho_{\tFD}\, U_\alpha^\dag
\eea
Here the $U_\alpha$ denote a large set of unitary transformations acting on the left times right Hilbert space that all commute both with the left and right Hamiltonian 
\bea
\label{ucomm}
[ \spc U_\alpha, H_L \spc ] \is [ \spc U_\alpha, H_R\spc ]  \, = \, 0.
\eea
We can think of equation \eqref{tmdsum} as the result of non-unitary time evolution of the thermo-field double state, which has undergone decoherence due to its interaction with a large environment. The decoherence is restricted by the condition that the energy on both sides remains identical at all times. Note that the above Ansatz has the property that the reduced density matrix on the {\small $L$} or {\small $R$} side Hilbert space looks exactly thermal. Indeed, we emphasize that all terms in the sum \eqref{tmdsum} are physically equivalent states: they differ only by the choice of the anti-unitary involution $\hat{T}$ used in defining the TFD state. Since there is no canonical choice for $\hat{T}$, all unitarily related TFD states are physically equivalent: they are mutually distinguishable but individually look identical. Hence it is natural to consider the decoherence process that produces an incoherent superposition of transformed TFD states.

If there are no exact degeneracies,  the condition \eqref{ucomm} implies that the $U_\alpha$ act by multiplying each energy eigen state with some phase $e^{i\alpha_i}$. If we further assume that the energy spectrum of the holographic CFT is completely random, and using that the
TFD state is annihilated by the $H_L\! -\nspc H_R$, we deduce that the $U_\alpha$ transformations can all be written as a time evolution operator
\bea
U_\alpha \is e^{-it_\alpha H_L}.
\eea
with $t_\alpha$ some long time typically of order the Poincar\'e recurrence time of the holographic CFT. If we further assume that the $t_\alpha$ represent a complete set of time instances, the sum \eqref{tmdsum} leads to complete dephasing. The resulting mixed state then looks like the old thermal mixed double state \eqref{latetmd} with only classical correlations between the two sides.

\pagebreak

To describe the new thermal mixed double state, we will now introduce two refinements (i)~we include the possibility that energy levels have a large (approximate) degeneracy  and (ii) we allow for the possibility that not all energy levels will undergo complete dephasing. Corresponding to these two refinements, we will distinguish three types of quantum numbers based on how they are correlated between the left- and right sectors {\small $L$} and {\small $R$}:

\addtolength{\parskip}{-1mm}
\begin{enumerate} 
\addtolength{\parskip}{-1mm}
\addtolength{\baselineskip}{-1mm}
\item{quantum numbers $j$ that are classically correlated between left and right}
\item{quantum numbers $m$ that are quantum entangled between left and right}
\item{quantum numbers $s$ that are classical and uncorrelated between left and right}\\[-7mm]
\end{enumerate}
\addtolength{\parskip}{1mm}

\noindent
Concretely, we will assume that the Hilbert space decomposes into sectors ${\cal H} = \Motimes_j {\cal H}_j$ labeled by $j$. We will call the $j$ quantum numbers  `Casimirs' and refer to the associated sectors as `modules' or `representations', since they are analogous to the Virasoro towers in 2D CFT. Each sector ${\cal H}_j$ is spanned by a set of basis states $|jms\rangle$ with energy
\bea
H |jms\rangle \is E_{jm}  |jms\rangle
\eea
dependent on $j$ and $m$ only. Hence the quantum number $s$ labels a set of (almost) exact degenerate energy eigen states. The  thermal state and TFD thus take the form
\bea
\label{rhothermal}
\  \rho(\beta) \!\is \!   \sum_{j,m,s} p_{jm} \, |jms\rangle \langle jms|, \\[3.75mm]
\  |\TFD\rangle \! \is\! \sum_{j,m,s} \sqrt{p_{jm}\!\nspc}\;\,  |jms\rangle_L | jms\rangle_R\\[3mm]
p_{jm} \spc = \spc & & \!\!\!\!\!\!\!\!\!\! \frac{e^{-\beta E_{jm}}}{Z(\beta)},  \qquad  Z(\beta) \spc =\nspc  \sum_{j,m,s} e^{-\beta E_{jm}}.
\eea

In the presence of (almost) exact degeneracies, the condition \eqref{ucomm} that $U_\alpha$ (almost) commutes with both Hamiltonians allows for more general solutions of the form
\bea
\label{uvee}
U_\alpha \is e^{-it_\alpha H_L} \otimes V_\alpha
\eea
where $V_\alpha$ denotes a unitary rotation acting on the $s$ quantum numbers only.
The action of the $U_\alpha$ in the sum \eqref{tmdsum} will thus in general decorrelate the $s$ quantum numbers on the {\small $L$} and {\small $R$} side. For generality and physical reasons, we further postulate that the $t_\alpha$ are all chosen such that the phases $e^{-it_\alpha E_{jm}}$ are independent of $m$. 
 The sum \eqref{tmdsum} then preserves coherence and perfect left-right entanglement of the $m$ quantum numbers. Physically, one can think of the $m$ quantum numbers as labeling the states in a protected code subspace within which all TFD states in the sum \eqref{tmdsum} look identical to the Unruh vacuum \cite{Verlinde:2012}-\cite{Almheiri:2015}.

To simplify the following discussion, let us consider the special case that all energy levels have the same degeneracy $e^{S_0}$ and that the energy $E_{jm}$ splits up in a $j$ dependent and $m$ dependent term 
\bea
s = 1, \ldots, e^{S_0} \qquad \qquad \ 
E_{jm} = \spc E_j + e_{m}\, .
\eea
The Hilbert space then factorizes into the tensor product of three sectors, one spanned with basis states labeled by $j$, one by $m$, and one by $s$. This tensor  product factorization will help make the properties of the three types of quantum numbers more evident.  We refer to the accompanying paper \cite{HV:2021} for a discussion of the more general case and for a direct path-integral derivation of the following explicit form of the thermo-mixed double state for a broad class of quantum systems.

\smallskip

With the above simplification, the thermal mixed double state factorizes into a tensor product of three density matrices, one for each type of quantum number
\bea
\label{tmdfacto}
\rho_{\tMD}\is \rho_{\mbox{\footnotesize $\!{\rm classical} \atop {\rm correlated}$}} \otimes\rho_{\mbox{\footnotesize \! ${\rm quantum} \atop {\rm entangled}$}}\otimes
\rho_{\mbox{\footnotesize \!\!${\rm classical} \atop {\rm uncorrelated}$}}
\eea
where\\[-8mm]
\bea
\label{tmdfactt}
  \rho_{\mbox{\footnotesize  \! ${\rm classical} \atop {\rm correlated}$}} \nspc \is \spc \sum_j \, p_j \, |\spc j\spc  \rangle_L \langle\spc j\spc| \spc \otimes \spc |\spc j\spc \rangle_R \langle\spc j\spc | \nonumber\\[3.5mm]
\rho_{\mbox{\footnotesize\!  ${\rm quantum} \atop {\rm entangled}$}} \; \is 
\sum_{m,m'} \sqrt{q_m q_{m'}\nspc} \, |m  \rangle_L \langle m'| \spc \otimes \spc |m \rangle_R \langle m' |\\[3.5mm]
\rho_{\mbox{\footnotesize\! \!${\rm classical} \atop {\rm uncorrelated}$}} \! \is \, e^{-2S_0} \, \sum_{s,s'} |s \rangle_L \langle s| \otimes |s' \rangle_R \langle s'| \nonumber
\eea
with\\[-7mm]
\bea
\label{tildezzz}
 p_j = \frac{e^{-\beta E_j}}{\tilde{Z}(\beta)}, \quad & &\quad \tilde{Z}(\beta) = \sum_j e^{-\beta E_j}\\[2mm]
  q_m =  \frac{e^{-\beta e_m}}{\zzz(\beta)},  \quad & & \quad  \zzz(\beta) = \sum_m e^{-\beta e_m}
 \eea
 We observe the three different types of correlations. The classical correlated factor looks like the old TMD state. The quantum entangled density matrix defines a pure thermo-field double state $\rho_{\raisebox{2pt}{\scriptsize\!  ${\rm quantum} \atop {\rm entangled}$}} = |{\rm tfd} \rangle\langle{\rm tfd}|$ within the $m$ sector. The third factor describes two decoupled maximally mixed states.
 
Figure 1 below gives a pictorial representation of the geometric entanglement and factorization properties of the new TMD state: the colored regions indicate how the three quantum numbers are shared between different segments of the double trumpet geometry associated with its second R\'enyi entropy
\vfill

\pagebreak

\begin{figure}[t]
\begin{center}

\includegraphics[scale=.73]{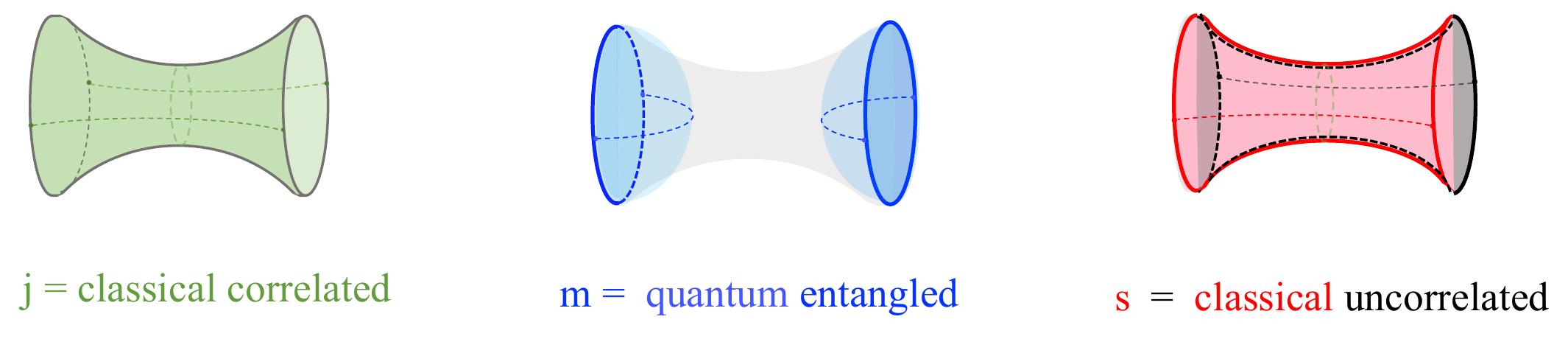}
\end{center}

\vspace{-1mm}
\caption{\addtolength{\baselineskip}{0.4mm} Depiction of the second R\'enyi entropy of the new thermo-mixed double state.\\[2mm]
The first factor in \eqref{tmdfacto} describes the classically correlated sector: the quantum number $j$ is shared both between the {\small $L$} and {\small $R$} sectors (correlated) as well as between the ket and bra states (classical). This sector in non-factorized, both between the two ends of the wormhole (left and right in the figure) and between the {\small $L$} and {\small $R$} sector (front and back).\\[2mm]
The second factor in \eqref{tmdfacto} describes the quantum entangled sector. This sector is in a pure state: the quantum number $m$ is shared coherently between the {\small $L$} and {\small $R$} sectors (entangled) but not across the ket and bra states (quantum). This sector in factorized between the two ends of the wormhole but the {\small $L$} and {\small $R$} sides (front and back) are connected.\\[2mm]
The third factor in \eqref{tmdfacto} describes the classical uncorrelated sector: the {\small $L$} and {\small $R$} sectors have independent quantum numbers $s$ and $s'$ (uncorrelated) but the ket and bra states share the same quantum number (classical). This sector in non-factorized between the two ends of the wormhole but the {\small $L$} and {\small $R$} sector (front and back) are disconnected.
}
\vspace{2mm}
\end{figure}

\bigskip

\section{Higher spectral form factor and R\'enyi entropy}
 \vspace{-1mm}
 
We now discuss the generalization of our replica Ansatz \eqref{zztmd} to higher order R\'enyi entropies.
For the time averaged quantities, this generalization is self-evident. We introduce higher order spectral form factors via (here $t_{ij} = t_i - t_j$)
\bea
\label{zzztfd}
{Z}(\beta\! + \! it_{12})\spc 
 \ldots  {Z}(\beta\! + \! it_{n1}) 
\is {Z(\beta)^n\strut} 
\,  {\rm tr}\bigl(\spc \rho_{\tFD}(t_1)\spc 
\ldots \rho_{\tFD}(t_n)\spc \bigr)\quad 
\eea
 As before, we expect that in the large time-differences limit this spectral form factor reaches a constant plateau value equal to 
its time average $\la Z(\beta)^n \ra  = \, Z(n\beta).$
The replica Ansatz 
\bea
\label{ansatznn}
 \tr(\rho_{\tMD}^n) \, = \, \frac{{\la Z(\beta)^n \ra}{\strut}}{Z(\beta)^n{\strut}}  
\eea
in the late time limit reduces to 
\bea
\label{oldrenyi}
\qquad \ \  \tr(\rho_{\tMD}^n) \, = \,\is \frac{ Z(n\beta)}{Z(\beta)^n} \, \sim \, e^{-(n-1) S_0}\qquad \qquad ({\rm plateau})
\eea
This equation is by construction satisfied by the old TMD state \eqref{latetmd}. Indeed, equation \eqref{oldrenyi} directly follows by taking the time average on both sides of equation \eqref{zzztfd}.
 
We would like to generalize the replica Ansatz \eqref{ansatznn} to the dip region,
where $\rho_{\tMD}$ now denotes the new generalized TMD state.
A concrete way to define both sides is to start from the expression \eqref{tmdsum} for the TMD state as a sum of unitary operators $U_\alpha$ acting on the TFD and then look for the combination of $U_\alpha$'s that minimizes the left-hand side of
\eqref{ansatznn}. As a second step, we then use the equality \eqref{zzztfd} to transfer this into a corresponding definition of the time and ensemble averaging on the right-hand side of \eqref{ansatznn}. With these definitions, the equality \eqref{ansatznn} automatically holds. Schematically, this procedure amounts to defining
\bea
\bigl\langle Z(\beta)^n \bigr\rangle
 & \!  \equiv\!&  \min_{t_{ij},\langle\ \rangle} \; \bigl\langle\spc {Z}(\beta\! + \! it_{12}) \, \ldots \, {Z}(\beta\! + \! it_{n1})\spc \bigr\rangle  \qquad \qquad  ({\rm near \ dip})
\eea
Here the minimization also applies to the averaging procedure as specified above.

It is reasonable to assume that the relationship between the dip and ramp behavior of the standard spectral form factor and the double trumpet partition function generalizes to higher orders, and that the leading contribution to the higher spectral form factor is provided by the partition function associated with the $n$-fold trumpet geometry
\bea
\qquad \bigl\langle Z(\beta)^n \bigr\rangle_{\! {}^{\mbox{\scriptsize{conn}}}} 
\!&\! \simeq\! &\! {Z(\Sigma_n)}\qquad \qquad \quad \Sigma_n \, = \,  \raisebox{-1.05cm}{$ \includegraphics[scale=.5]{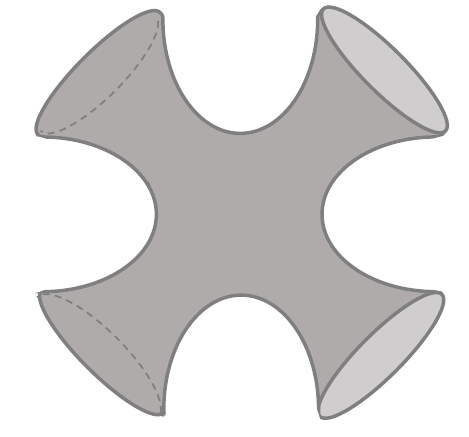}$}.
\eea
This equality makes a concrete prediction for the $e^{S_0}$ scaling of the $n$-th R\'enyi entropy
\bea
\label{nrenscale}
\quad\; \tr(\rho_{\tMD}^n) \! & \! 
= \! &  \! \frac{ {Z(\Sigma_n)}}{Z(D)^n} \,  \sim \, e^{-2(n-1) S_0}  \qquad \qquad  ({\rm near \ dip})
\eea
The power of $e^{S_0}$ equals the difference 
between the Euler character of the $n$-fold trumpet $\Sigma_n$ and $n$ separate disks $D$.
Again, this scaling exponent is twice as big as for the R\'enyi entropy of the old TMD state. However, the scaling \eqref{nrenscale} is matched by the $n$-th R\'enyi entropy of the new TMD state \eqref{tmdfacto}-\eqref{tmdfactt} 
\bea
\label{tmdnrenyi}
\tr(\rho_{\tMD}^n)\! \is\! e^{-2(n-1)S_0} \sum_j p_j^n\spc=\spc e^{-2(n-1)S_0} \, \frac{\tilde{Z}(n\beta)}{\tilde{Z}(\beta)^n}\, 
\eea
with $\tilde{Z}(\beta)$ defined in \eqref{tildezzz}. This match supports our proposal that the new TMD Ansatz has the right quantum statistical properties to describe the dip region.

It is instructive to deconstruct the expression \eqref{tmdnrenyi} into a rule for computing expectation values in the ensemble averaged system. Let us decompose the partition function into a sum over subspace-partition functions with given value of the Casimir quantum number $j$
\bea
\label{zjsum}
Z(\beta) \! \is \! \sum_j Z_j(\beta)
\eea
Equating \eqref{tmdnrenyi} and \eqref{ansatznn} indicate that we should adopt the following rule for the expectation values of the subspace-partition functions (c.f. \cite{DonHenry})
\bea
\bigl\langle Z_{j_1}(\beta) Z_{j_2}(\beta) \ldots Z_{j_n}(\beta) \bigr\rangle_{\rm conn}\! \is \! \left\{\begin{array}{c} e^{-(2n-2) S_0} 
{\raisebox{0pt}{$Z_j(\beta)^n$}_{\spc} }
\qquad {\rm if}\ \ j_i
\nspc = \nspc j \quad \forall i \\[2.5mm]
\qquad {0 \qquad\qquad \qquad \, {\rm otherwise}} \end{array} \right.
\eea
This formula combines two combinatorial rules (i) the subsystem partition functions act like projection operators onto orthogonal subspaces and (ii) the connected expectation value is suppressed by a factor of $e^{-S_0}$ for each additional insertion of $Z_j(\beta)$. One the holographic side, $Z_j(\beta)$ represents a loop operator that creates a hole with a given $j$ sector running around the loop. Each extra hole reduces the partition function by a factor $e^{-S_0}$. The $j$ quantum number is shared across the $n$-fold trumpet geometry connecting the $n$ holes \cite{HV:2021}. 

With these rules, we evaluate
\bea
\la Z(\beta)^n  \ra\is e^{-(2n-2) S_0} \sum_j \,
{\raisebox{0pt}{$Z_j(\beta)^n$}_{\spc} }
\eea
For our simplified example with the factorized spectrum, we have $Z_j(\beta) = \spc e^{S_0} \spc e^{\mbox{\scriptsize$-\beta E_j$}}\spc \zzz(\beta)$. The above formula then gives
\bea
\label{zzzn}
\la Z(\beta)^n\ra \is e^{-(n-2) S_0} \, \tilde{Z}(n\beta)\, \zzz(\beta)^n
\eea
Via \eqref{ansatznn}, this reproduces the result \eqref{tmdnrenyi} for the R\'enyi entropy.  
The result \eqref{zzzn} is verified via a direct path integral computation for a class of quantum systems in \cite{HV:2021}.

The geometric entanglement and factorization properties across $\Sigma_n$  of the three types of quantum numbers $j,m,s$ is the same as discussed before and indicated in the figure below.${\strut}$

\begin{center}
\includegraphics[scale=.64]{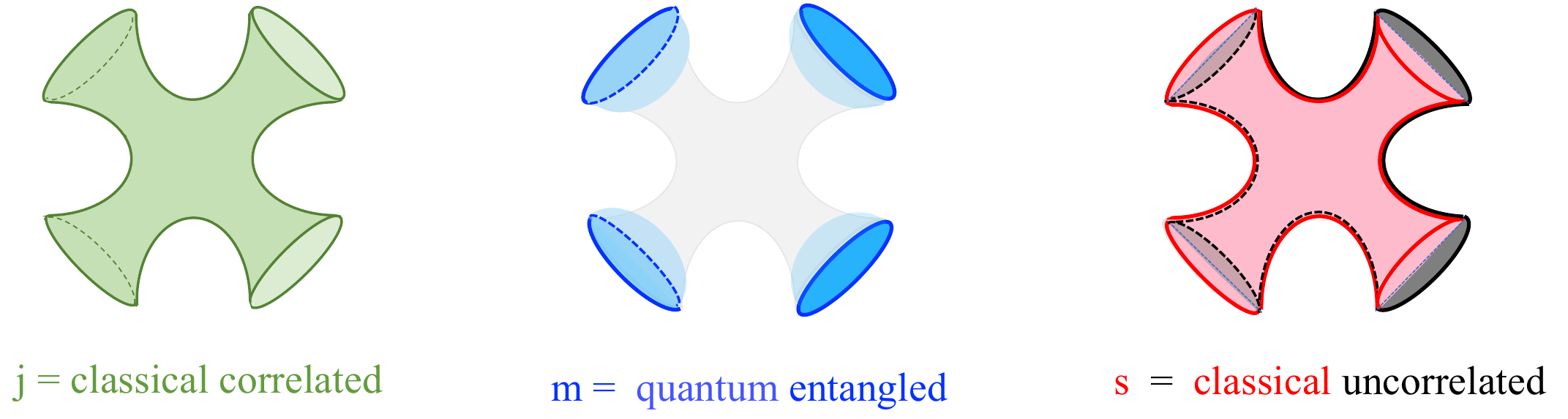}
\end{center}

\bigskip

\section{Von Neumann entropy and mutual information}
\vspace{-.3mm}

It is instructive to compare the von Neumann entropy and mutual information carried by the various thermal mixed states in our story. We specialize to our simplified model system with the facorized spectrum. The von Neumann entropy $S(\beta) = - \tr(\rho(\beta) \log \rho(\beta))$ of the standard thermal state \eqref{rhothermal} thus slits into three terms 
\bea
& & \qquad \qquad \qquad \qquad\qquad \qquad {S_{\raisebox{1pt}{\ppp}}(\beta) =- \sum_j p_j \log p_j }\nonumber \\[-3mm]
S(\beta) \is  \, S_{0}  
+ S_{\raisebox{1pt}{\ppp}} (\beta)
+ S_{\raisebox{1pt}{\qqq}}(\beta)\qquad \qquad \qquad \\[-3mm]
& & \qquad \qquad \qquad \qquad \qquad \qquad {S_{q}(\beta)\spc = \spc - \sum_m q_{m} \log q_{m} }\nonumber
\eea
with $p_j$ and $q_m$ the respective Boltzmann weights of the $j$ and $m$ quantum numbers. The $S_0$ term accounts for the contribution of the overall degeneracy of the energy levels. 

Let us compare three types of mixed and/or entangled states of the two-sided system (i)~the factorized thermal state  $\rho_L \! \otimes \nspc \rho_R$, (ii) the pure thermo-field double state $\rho_{\tFD}$ and (iii)~the new thermo-mixed double state $\rho_{\tMD}$. For all three states, the reduced density matrix of each side is thermal. This means that the sum of the von Neumann entropy $S_{LR}$ of the combined state and the mutual information $I_{LR} = S_L + S_R - S_{LR}$ always adds up to  twice the thermal entropy of the one sided system: $S_{LR} + I_{LR} = 2S(\beta)$. The three thermal states are distinguished by how this total entropy $2S(\beta)$ is split up between the two terms.

Evidently, the factorized thermal state has twice the thermal entropy of the single sided system and carries no mutual information between the two sides
\bea
S(\rho_L \! \otimes \nspc \rho_R)\!  \nspc\is\!  2S_0 \nspc + \nspc 2 S_{\raisebox{1pt}{\ppp}}(\beta)\nspc + \nspc 2 S_{\raisebox{1pt}{\qqq}}(\beta), \qquad \qquad\ \ \ I_{LR}(\rho_L\! \otimes\nspc \rho_R)\smpc = \smpc 0.
\eea
The thermo-field double state, on the other hand, has zero von Neumann entropy and carries maximal mutual information between $L$ and $R$
\bea
S(\rho_{\tFD})\spc \is\spc  0, \qquad \qquad\ \ I_{LR} (\rho_{\tFD})\nspc = \nspc 2S_0 \nspc  + \nspc 2 S_{\raisebox{1pt}{\ppp}}(\beta)\nspc  +\nspc 2 S_{\raisebox{1pt}{\qqq}} (\beta).
\eea
The thermo-mixed double splits the difference between the two: the classically correlated quantum number $j$ contributes an equal amount of entropy and mutual information, the entangled quantum number $m$ does not carry entropy and only contributes to the mutual information, and the uncorrelated $s$ quantum number only contributes to the entropy and carries no mutual information:
\bea
S(\rho_{\tMD})  \is\nspc  2S_0 + S_{\raisebox{1pt}{\ppp}}(\beta) \spc , \qquad \qquad\ \  I_{LR}(\rho_{\tMD}) \spc = \spc  S_{\raisebox{1pt}{\ppp}}(\beta)  + 2 S_{\raisebox{1pt}{\qqq}}(\beta)  .
\eea
Hence we can consider several different limits of this TMD state, depending on which of the three quantum numbers contributes the most to the spectral density.  
\bigskip

\def\VV{\mbox{$\mathcal{V}$}}

\section{Conclusion: non-factorization and decoherence}
\vspace{-.3mm}

We have formulated a link between wormhole saddle points that contribute to the spectral form factor and replica wormholes that appear in the computation of the Rényi entropy of a radiating two-sided black hole. In the first case, the wormholes contribute  non-factorized terms that quantify the destructive interference between the partition sums of two decoupled systems, while in the latter case, the wormholes capture the decoherence and purification of black hole radiation and the degrading of the entanglement across the ER bridge. Hence our story provides a link between non-factorization, entanglement and decoherence. 

The key element in our story is  the replica Ansatz \eqref{zztmd} for the spectral form factor. It equates the leading order non-factorized contribution $\langle Z(\beta)^2\rangle$ to the spectral form factor to the second R\'enyi entropy, or purity, of a suitable mixed state $\rho_{\tMD}$ of the two sided system. We argued that this thermo-mixed double state $\rho_{\tMD}$ takes the general form an incoherent sum  \eqref{tmdsum} of transformed thermo-field double states. The physical justification for this Ansatz is two-fold. First, all transformed TFD states $U_\alpha\spc \rho_{\tFD} \spc U_\alpha^\dag$ that contribute in \eqref{tmdsum} are physically equivalent and have equal right to be called the TFD state. Since they all have the same macroscopic properties it is natural to consider the statistical superposition of all such states. 
 Second, in the accompanying paper \cite{HV:2021} we will show, via an explicit path-integral computation and for a general class of quantum systems, that the $n$-th R\'enyi entropy  of the new TMD state indeed matches with the geometric partition function $Z(\Sigma_n)$ associated with the $n$-fold replica wormhole geometry $\Sigma_n$. 

Plugging the expression \eqref{tmdsum} into the replica Ansatz \eqref{zztmd} yields the following definition for the averaged spectral form factor
\bea
\label{zzsum}
\la Z(\beta)^2 \ra \! \is\! \frac{1}{N} \sum_{\alpha} \, \bst\spc \tr\lbl e^{-\beta H} U_{\alpha}\spc  \rbl\bst^2 
=\spc \frac{1}{N} \sum_{\alpha} \, \bst\spc \tr\lbl e^{-(\beta+it_{\alpha})H} V_\alpha\spc 
\rbl\bst^2. 
\eea
Hence in our proposed definition, the averaging does not involve a sum over different quantum systems but rather a sum over different time instances $t_\alpha$ and unitary rotations $V_\alpha$ acting on the degenerate energy eigen spaces.\footnote{Of course, we can view \eqref{zzsum} as a sum over Hamiltonians $H_\alpha$ specified by $e^{-(\beta+it)H_\alpha} = e^{-(\beta+it_\alpha) H} \otimes V_\alpha$.} The sum over time instances leads to dephasing that eliminates the off-diagonal matrix elements in $\rho_{\tFD}$ and replaces the entanglement between the {\small $L$} and ${\small R}$ sectors by classical correlation. The unitary rotation $V_\alpha$ can be thought of as the insertion of a topological interface between the {\small $L$} and ${\small R}$ sector, and the sum over $V_\alpha$ can thus be viewed as a sum over such topological interfaces. This sum decorrelates the quantum numbers $s$ that label the degenerate energy levels. Finally, there may in general be a code subspace that does not dephase; its contribution to the spectral form factor remains factorized. From the point of view of the TFD states, this sector remains entangled.

 The explicit formula  \eqref{tmdfacto}-\eqref{tmdfactt} for the TMD state
 can be rewritten as an incoherent sum of partially dephased TFD states
\bea
\label{tmdnew}
\rho^{\rm new}_{\tMD}\! \is  \!  e^{-2S_0} \sum_{j, s,s'}\; {p_{j}}\;|\spc \TFD_j\spc \rangle_{\nspc ss'}\smpc \langle \spc \TFD_{j}\spc |_{ss'} 
\eea
where $ |\TFD_j\rangle_{ss'}$ denotes the TFD state restricted to a $j$ eigen sector,  labeled by two uncorrelated quantum numbers $s$ and $s'$
\bea 
\label{tmdss}
  |\spc \TFD_j\spc \rangle_{ss'} \! \is\!  \sum_m \, \sqrt{q_{m}\!} \;  | jms\rangle_{\nspc L}\,  | jms'\rangle_{\nspc R}  
\eea
Here $p_j$ and $q_{m}$ denote the respective Boltzmann weights. The $m$ quantum number is in a pure entangled state: to operators that only act on this `code subspace', the state \eqref{tmdnew} looks like the standard thermo-field double state. So if we would assume that this code subspace represents the space of low energy observables of a bulk observer, the  bulk geometry described by \eqref{tmdnew} would look like a two-sided black hole with a smooth ER bridge.

In \eqref{tmdnew} we put the superscript `new' to distinguish this state from the fully dephased `old' thermo-mixed double state introduced in \cite{HV:2020}\cite{Takayanagi:2020} and in equation \eqref{latetmd}. In the $jms$ notation, the old TMD state takes the form
\bea
\label{tmdold}
\rho^{\rm old}_{\tMD}\! \is  \!  e^{-S_0} \sum_{j, s}\; {p_{j}}\;|\spc \TFD_j\spc \rangle_{ss} \smpc \langle \spc \TFD_{j}\spc |_{ss} .
\eea
Here we placed the $s$ quantum numbers in a dephased classically correlated state, but chose to still preserve the entanglement in the code subspace spanned by the $m$ quantum numbers. Plugging the new and old TMD into the replica Ansatz gives 
\bea
\qquad \frac{\langle Z(\beta)^n\rangle}{ Z(\beta)^n}\nspc \is \nspc \tr\lbl(\rho^{\rm new}_{\tMD})^n\rbl \spc = \spc e^{-2(n-1)S_0}\sum_{j} \,  p_j^n\quad \qquad \mbox{(near dip)}\\[3mm]
\qquad \frac{\langle Z(\beta)^n\rangle}{ Z(\beta)^n} \nspc \is \nspc \tr\lbl(\rho^{\rm old}_{\tMD})^n\rbl \spc = \spc e^{-(n-1)S_0}\sum_{j} \, p_j^n \quad \qquad\ \mbox{(plateau)}
\eea
Both formulas seem to capture a universal type of behavior. Which of the two dominates depends on the physical question and the time scales involved.
Holographic quantum systems typically have a chaotic spectrum without exact degeneracies, but they may have a large number of approximately degenerate energy states \cite{Kitaev}\cite{Maldacena:2016}. In this situation one would expect that at very long time scales, the system will eventually relax to the old TMD state. 

This then raises the question: which among all the above universal thermal states describes a generic two-sided black hole geometry? The proposal put forward in \cite{HV:2020} is that, unless the system is fine tuned, a typical two-sided black hole is best described by the old TMD state. From the quantum statistical perspective, this looks  reasonable. Understanding this from the semi-classical gravity side, however, remains an open problem.

\section*{Acknowledgements}
\vspace{-2mm}

It is a pleasure to thank Akash Goel, Clifford Johnson, Joaquin Turiaci, Erik Verlinde, Mengyang Zhang and Wenli Zhao for helpful discussions and comments. This research is supported by NSF grant number PHY-1914860. 


\end{document}